\documentclass[twocolumn,showpacs,superscriptaddress]{revtex4-1}
\usepackage{amsbsy}
\usepackage{amsfonts}
\usepackage{amssymb}
\usepackage{amsmath}
\usepackage{graphicx,color,subfigure,pstricks}

\newcommand{\ie}{{\it{i.e.}}}
\newcommand{\etal}{{\it{et al.}}}

\newcommand{\Tr}{\mathrm{ Tr }}

\newcommand{\identity}{\openone}
\newcommand{\de}[1]{\left( #1 \right)}
\renewcommand{\Re}[1]{{\mathrm{Re}}\de{#1}}

\begin{document}

\title{Intensive temperature and quantum correlations for refined quantum measurements}

\author{Alessandro Ferraro}
\affiliation{ICFO-Institut de Ciencies Fotoniques, Mediterranean Technology Park, 08860 Castelldefels (Barcelona), Spain}
\affiliation{Department of Physics and Astronomy, University College London,  London WC1E 6BT, United Kingdom}
\author{Artur Garc\'ia-Saez}
\affiliation{ICFO-Institut de Ciencies Fotoniques, Mediterranean Technology Park, 08860 Castelldefels (Barcelona), Spain}
\affiliation{Departament d'Estructura i Constituents de la Mat\`{e}ria, Universitat de Barcelona, 08028 Spain}
\author{Antonio Ac\'in}
\affiliation{ICFO-Institut de Ciencies Fotoniques, Mediterranean Technology Park, 08860 Castelldefels (Barcelona), Spain}
\affiliation{ICREA-Instituci\'o Catalana de Recerca i Estudis Avan\c cats, Lluis Companys 23, 08010 Barcelona, Spain}

\pacs{03.67.-a, 05.30.-d, 05.70.-a}

\begin{abstract}We consider the concept of temperature in a setting
beyond the standard thermodynamics prescriptions. Namely, rather
than restricting to standard coarse-grained measurements, we
consider observers able to master any possible quantum
measurement---a scenario that might be relevant at nanoscopic
scales. In this setting, we focus on quantum systems of coupled
harmonic oscillators and study the question of whether the
temperature is an intensive quantity, in the sense that a block of
a thermal state can be approximated by an effective thermal state
at the same temperature as the whole system. Using the quantum
fidelity as figure of merit, we identify instances in which this
approximation is not valid, as the block state and the reference
thermal state are distinguishable for refined measurements.
Actually, there are situation in which this distinguishability
even increases with the block size. However, we also show that the
two states do become less distinguishable with the block size for
coarse-grained measurements ---thus recovering the standard
picture. We then go further and construct an effective thermal
state which provides a good approximation of the block state for
any observables and sizes. Finally, we point out the role
entanglement plays in this scenario by showing that, in general,
the thermodynamic paradigm of local intensive temperature applies
whenever entanglement is not present in the system.
\end{abstract}

\maketitle A characteristic trait of macroscopic matter is the
simplicity with which it can be typically described in physical
terms. The long-lasting success of thermodynamics is built upon
this evidence: few thermodynamic variables are sufficient to
effectively describe a piece of matter made of a huge number of
particles. The origin of this simplification lies in that
macroscopic objects are usually probed by extremely {\it coarse
measurements}. Specifically, macroscopic observations sense only
averages and just few properties---namely, thermodynamic variables
like entropy or temperature---suffice to describe the system after
such averaging \cite{Cal}.

A question arises whether this picture breaks down when the
standard requirement of coarse measurements is relaxed and more
general measurements---in particular, more {\it refined
measurements}---are at disposal. Of course, the use of
refined measurements goes beyond the standard thermodynamic
prescriptions. Nonetheless, at nanoscopic scales, these
measurements are foreseeable \cite{nanoth} and this may imply
significant deviations from the standard (macroscopic and coarse
grained) thermodynamic scenario. In particular we will consider
here quantum systems for which some deviations from thermodynamics
have been already explored at small scales \cite{ANFO,HMH}.
Furthermore, a series of recent results suggest that some
hypothesis commonly invoked in thermodynamics can be actually
relaxed in a quantum setting. This is for example the case of
subjective lack of knowledge---usually invoked to prove the
emergence of the canonical ensemble---which has been shown to be
unnecessary \cite{typic}. In general there are evidences that
thermodynamics principles can be applied to non-standard
scenarios, beyond the ones originally envisaged \cite{GMM}.

Here we address these considerations focusing on the concept of
temperature. In particular, we test whether the temperature is an
intensive quantity by taking into account both coarse- and
refined-measurement scenarios. In fact, considering a quantum
setting, previous results have shown that the temperature may not
be intensive at small scales---where coarse measurements more
evidently show their inadequacy---resulting in the fact that
subparts of thermal states may no longer be described as thermal
states with the same global temperature as the whole system
\cite{GMM}. However the role of measurements in this context has
not been taken into account in detail. In addition, the mechanism
that originates this departure from the standard thermodynamic
behaviour is still unclear. In analogy with the case of classical
systems, one might expect that the energy balance between subparts
of the system should be the main responsible for it. However, we
show that this is not the case and that other mechanisms play a
major role.

In order to tackle the foregoing questions, we consider quantum
systems consisting of coupled harmonic oscillators in a thermal
state \cite{note_thermalization}. We study the distinguishability
between a block of harmonic oscillators and a reference thermal
block at the same temperature as the whole system. We first
provide instances in which the temperature is no longer intensive,
in the sense that under refined measurements the state of the
block can be distinguished from the reference thermal state.
Contrary to the intuition stemming from the standard setting of
coarse-grained measurements, the breakdown of intensiveness is
more easily observed for larger systems. Second, despite this
first result, we see how the standard thermodynamic situation
(where the state of the block and the reference thermal state do
become more indistinguishable for larger systems) is recovered for
coarse-grained observables. Third, we show that for any type of
observables it is possible to define an appropriate effective
thermal state approximating the block state, even for small sizes.

As said, the origin of the deviation from intensiveness relies on
the capacity to perform refined measurements. The fact that larger
systems exhibit larger deviations reveals that the energy balance
between subpart of the system does not play a significant role. On
the other hand we will relate this deviation with the presence of
correlations in the system. Specifically, we show that
the presence of genuine quantum correlations is related to the
departure from the intensive behaviour: when quantum entanglement
is significantly present, the temperature ceases to be intensive
for refined measurements; vice-versa, for vanishing entanglement
the thermodynamic paradigm of intensive temperature applies for
\textit{any} possible measurements.

Our approach to this problem stems from concepts and methods
proper of quantum information theory \cite{NC}, for a twofold
reason. First, as said, we assess the intensiveness of temperature
by considering the distinguishability between a region of the
system and a thermal state. For this purpose we use the quantum
fidelity, a quantum information concept that quantifies the
distinguishability between two states given \textit{any} possible
measurement. Second, in order to investigate the origin of the
non-intensive behaviour of temperature, we distinguish quantum
from classical correlations. This requires the use of entanglement
quantifiers, since correlation functions commonly considered in
condensed matter theory are not sensitive to this distinction.
\section{Local states and temperature}
Let us consider a system $S$ composed of a macroscopic number of
elementary constituents that---after having thermalized with a
proper environment---lies in a canonical state. The interactions
among the system are described by the Hamiltonian $H$.
Following Refs.~\cite{HMH,GMM,GFA}, we adopt here the following
notion of {\em local intensiveness}: given $S$ at temperature $T$,
we say that the temperature is locally intensive when the block
$B$ can be described by a canonical state at the same temperature
$T$. More specifically, the system state is given by
$\Omega(\beta)=\exp[-\beta H]/Z$, where $\beta$ is the inverse
temperature, and $Z$ the partition function.  The actual state of
the block is thus given by
\begin{equation}\label{block_state}
    \rho_B(\beta)=\Tr_R\Omega(\beta)=\Tr_R(\exp[-\beta H]/Z) ,
\end{equation}
where the trace is over the rest of the system. We say that the
temperature is intensive when
$\rho_B(\beta)\approx\Omega_B'(\beta)$, being $\Omega_B'(\beta)$ a
thermal state for the block,
\begin{equation}\label{block_thermal}
    \Omega_B'(\beta)=\exp[-\beta H_B']/Z',
\end{equation}
where $H_B'$ is an effective Hamiltonian acting only on $B$.

Clearly, $H_B'$ cannot be left arbitrary. Here we introduce a
generic procedure to identify a proper block Hamiltonian. In
particular, we impose the following requirements to $H_B'$:
$\rm{(R_1)}$ it is temperature independent; $\rm{(R_2)}$ it gives
rise to an intensive behaviour for high temperatures. Requirement
($R_1$) is motivated by the fact that if $H_B'$ was free to change
with $\beta$ the problem would lose relevance. In fact, any state
$\rho$ can be trivially written as $\rho=e^{-\beta H}/Z$ with
arbitrary $\beta$ for a proper $H$. Concerning ($R_2$), it is
motivated by the physical request of recovering the standard
intensive behaviour for high temperatures. The choice of these two
requirements, a part from the mentioned physical motivations,
turns out to have sensible benefits \textit{a posteriori}. In
fact, we will see that this procedure singles out a unique
Hamiltonian $H'_B$ which, in turn, coincides with the quantized
version of the classical Hamiltonian associated to the classical
analogue of our quantum system. Let us stress that here we
consider these two requirements for a specific family of
Hamiltonian systems. However, the procedure can be applied more in
general (see also Ref.~\cite{GFA}).

In what follows, we quantify the degree of temperature intensiveness by the distinguishability between the two quantum states, $\rho_B(\beta)$ and $\Omega_B'(\beta)$, under any possible measurement. This concept is captured by the quantum fidelity \cite{Fidelity}, defined for two generic states $\sigma_1$ and $\sigma_2$ as $F[\sigma_1,\sigma_2]={\rm Tr}\left[\sqrt{\sigma_1}\sigma_2\sqrt{\sigma_1}\right]^{1/2}$. Recall that the fidelity is one only when two states are identical, and a given amount of fidelity gives the distance between the two set of data obtained from the measurement that ---among all possible measurements--- distinguishes at best the two considered states. We introduce then the intensive fidelity
\begin{equation}\label{intfid}
F_I(\beta)=F\left[\rho_B(\beta),\Omega_B'(\beta)\right]\;,
\end{equation}
which decreases as temperature loses its intensive character and approaches one when it is intensive. In the latter case, no measurement---no matter how refined---can distinguish the state of the block from a thermal one.

From now on, we will say that the temperature undergoes a
breakdown of intensiveness whenever $F_I(\beta)$ is significantly
lower than one. As said this means that an observer able to
perform refined measurements can detect that the actual
state of block $B$  is not a thermal state at the same temperature
as the whole system $S$. However, this by no means implies that
the temperature ceases to be intensive for standard thermodynamic
measurements---an undisputed property that clearly remains valid
under the prescriptions of thermodynamics. Let us also stress here
that the fidelity is a very sensitive measure, in the sense that
apparently small differences between two states can lead to very
low values of the fidelity \cite{note_Fsensitivity}. This,
rather than being a misbehaviour, is just a consequence of the
fact that fidelity quantifies the distinguishability of two states
under any possible measurement.
\section{Harmonic systems}
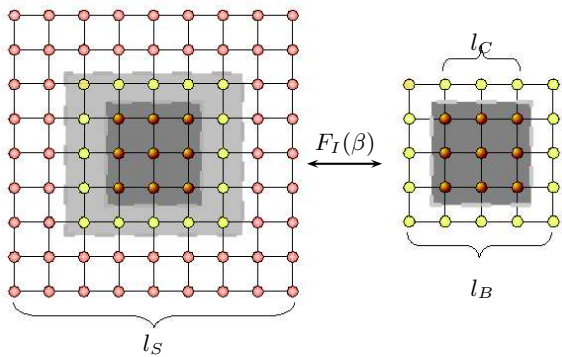
\begin{figure}
\includegraphics[width=8cm]{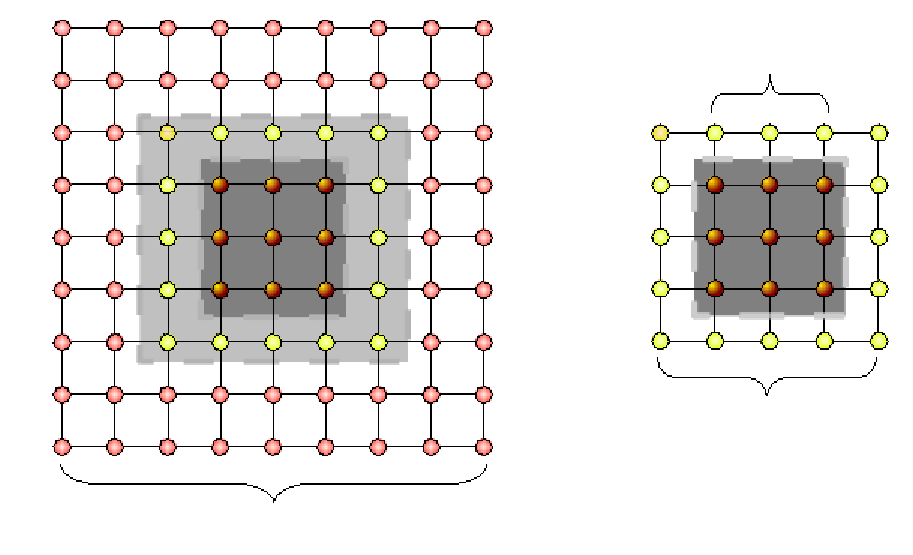}
\begin{picture}(3,3)
\put(-167,4){$l_S$}
\put(-43,25){$l_B$}
\put(-43,118){$l_C$}
\put(-102,80){$F_I(\beta)$}
\put(-105,75){\psline{<->}(0,0)(1,0)}
\end{picture}
\caption{Schematic representation of a two-dimensional harmonic
lattice composed of $n_S=l_S\times l_S$ oscillators set up in a
thermal state. The reduced state of a distinguished block of it
(light plus dark grey region), composed of $n_B=l_B\times l_B$
oscillators, is compared to a thermal state of a $l_B\times
l_B$-oscillator system, as in the right, via the fidelity
$F_I(\beta)$ (see text). The cores of the two systems, both
composed of $n_C=l_C\times l_C$ oscillators (dark grey regions),
are also shown.}\label{lattice}
\end{figure}
In this work we focus on quantum harmonic systems (see
Fig.~\ref{lattice}) composed of $n_S$ interacting oscillators
defined by position and momentum operators $q_i$ and $p_i$
respectively ($i=1,\dots,n_S$). Introducing the vector
$X=(q_1,\dots,q_{n_S},p_1,\dots,p_{n_S})$, the Hamiltonian
$H=X(V\oplus\identity_S)X^T$
describes the system, with $\identity_S$ denoting the $n_S\times
n_S$ identity matrix. Here the ground and thermal states are
Gaussian, thus permitting the use of powerful methods---developed
in the context of quantum information theory---for the calculation
of fidelity \cite{PS} and quantum correlations \cite{Gaussian}.
Hamiltonian $H$ models a variety of physical systems, ranging from
vibrational degrees of freedom in crystal lattices and ion traps
to free scalar Klein-Gordon field on a lattice \cite{ECP}.
Besides, being exactly solvable \cite{note_integrable}, these
models are a standard testbed in quantum thermodynamics
\cite{ANFO,HMH,GMM}.

We first derive the effective Hamiltonian $H_B'$ for the block by
imposing the requirements $\rm{(R_1)}$ and $\rm{(R_2)}$ introduced
above. Being Gaussian, the states $\Omega(\beta)$ are completely
described by their covariance matrix $\gamma(\beta)$. The elements
of the latter are defined as
$\gamma(\beta)_{kl}=\Re{\Tr\{\Omega(\beta)[X_k-\bar X_k][X_l-\bar
X_l]\}}$, where $\bar X_k=\Tr\{\Omega(\beta)X_k\}$. The explicit
expression of $\gamma(\beta)$ is
\begin{equation}
\gamma(\beta)=[V^{-1/2}W(\beta)]\oplus[V^{1/2}W(\beta)]\,,
\end{equation}
where $W(\beta)=\identity_S +2[\exp(\beta V^{1/2})
-\identity_S]^{-1}$. The high temperature limit of
$\gamma(\beta\rightarrow 0)$ is given by
$\frac{2}{\beta}[V^{-1}\oplus\identity_S]$. From the latter one
can easily obtain the high temperature limit of the block
covariance matrix, denoted as $\gamma_B(\beta\rightarrow 0)$. Let
us recast $V$ as a block matrix
\begin{equation}\label{blockV}
V=
\left(
\begin{array}{cc}
V_B & V_{BR} \\
V_{BR}^T & V_R
\end{array}
\right)\;,
\end{equation}
where $V_{B}$~($V_{R}$) refers to the block (rest) and $V_{BR}$ to
the interaction between $B$ and $R$. In this notation
$\gamma_B(\beta\rightarrow
0)=\frac{2}{\beta}[V'^{-1}\oplus\identity_B]$, where we introduce
an effective potential matrix $V'=V_B-V_{BR}V_R^{-1}V_{BR}^T$. The
unique choice of the effective Hamiltonian $H_B'$ of the block is
$H_B'=X_B(V'\oplus\identity_B)X_B^T$, where $X_B$ denotes the
operators referring to the block, which satisfies both
requirements (R1) and (R2).

Before proceeding further, let us consider the classical analogue
of the quantum systems under consideration. This helps in
clarifying our approach for the choice of $H_B'$ and in singling
out the genuine quantum aspects responsible for the intensiveness
breakdown. Let us denote with $H_c$ the classical version of the
Hamiltonian $H$, where the positions and momenta in $X$ are
classical phase space coordinates. At thermal equilibrium, the
system is described by a Boltzmann-Gibbs distribution
$P(X)=e^{-\beta H_c}/Z$. The distribution of the block $P_B(X_B)$
is obtained after integrating $P(X)$ over the oscillators in $R$.
This results in the distribution $P_B(X_B)=e^{-\beta
H'_{B,c}}/Z'$, where $H'_{B,c}=X_B(V'\oplus\identity_B)X_B^T$ and
the potential matrix turns out to coincide with $V'$ introduced
above. Indeed, this further clarifies the physical meaning of
$H'_B$, since it can be interpreted as the quantized version of
$H'_{B,c}$. Notice that since $V'\neq V_B$ the oscillators of the
block are subjected to a renormalization of their bare frequency,
a well known effect in open systems \cite{Wei}. Remarkably, the
expression of $P_B(X_B)$ shows that the temperature is always
intensive in classical harmonic systems (for any block size,
temperature, and coupling). This fact already suggests that any
possible deviation from intensiveness in quantum harmonic systems
should have a genuine quantum origin.

In what follows, we focus on one- and two-dimensional harmonic
lattices endowed with periodic boundary conditions and composed of
$n_S$ oscillators interacting with their nearest neighbours. The
potential matrix for 1D systems can be expressed as a $n_S \times
n_S$ circulant matrix of the form $V_1={\rm
circ}\{1,-c,0,...0,-c\}$. For 2D systems, composed of $n_S = l_S
\times l_S$  oscillators (see Fig.~\ref{lattice}), one has instead
a block-circulant matrix:
\begin{equation}
V_2={\rm circ}\{V_1,-c \identity_{l_S},0_{l_S},...,0_{l_S},-c \identity_{l_S}\}\;.
\end{equation}
where $V_1$, $\identity_{l_S}$, and $0_{l_S}$ are in turn
$l_S\times l_S$ matrices. The coupling parameter $c$ belongs to
$[0,1/2^d)$, where $d=1,2$ is the dimension, and the system is
critical at zero temperature for $c\rightarrow 1/2^d$.

\section{Intensiveness breakdown for refined quantum measurements}
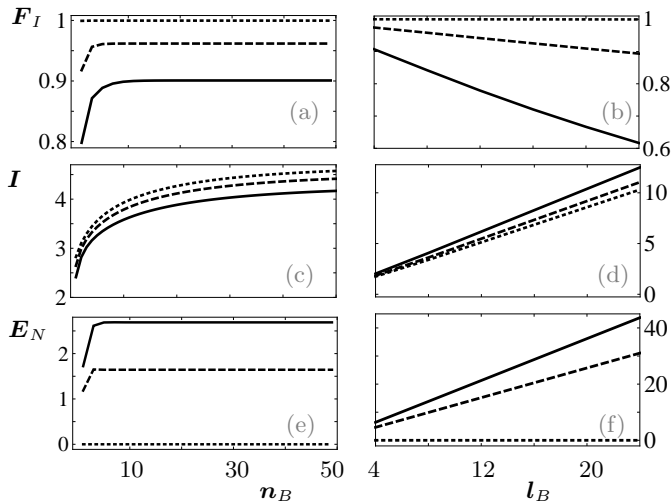
\begin{figure}
\subfigure{\label{F1_nB}
\includegraphics[width=4.1cm]{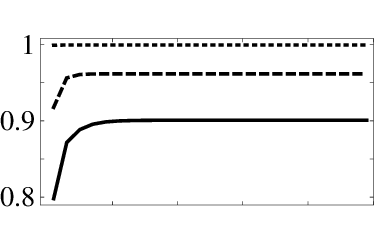}}
\subfigure{\label{F2_LB}
\includegraphics[width=4.1cm]{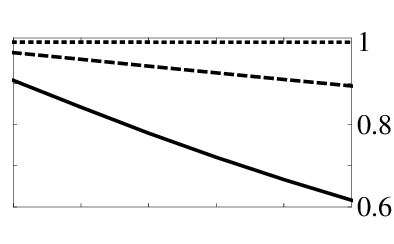}}
\begin{picture}(0,0)(133,110)
\put(-120,168){$\boldsymbol F_I$}
\put(-15,132){\gray (a)}
\put(103,132){\gray (b)}
\end{picture}\\
\vskip -0.8cm
\hskip 0.16cm
\subfigure{\label{MI1_nB}
\includegraphics[width=3.9cm]{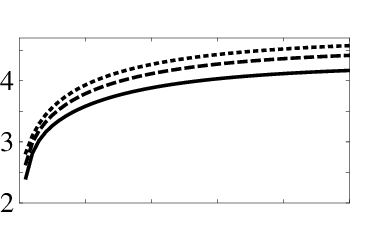}}
\subfigure{\label{MI2_LB}
\includegraphics[width=4.03cm]{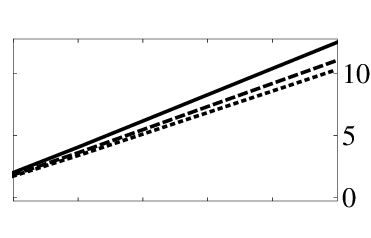}}
\begin{picture}(0,0)(133,115)
\put(-120,168){$\boldsymbol I$}
\put(-15,132){\gray (c)}
\put(103,132){\gray (d)}
\end{picture}\\
\vskip -0.8cm
\hskip 0.22cm
\subfigure{\label{EN1_nB}
\includegraphics[width=3.85cm]{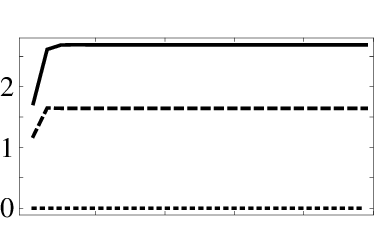}}
\subfigure{\label{EN2_LB}
\includegraphics[width=4.05cm]{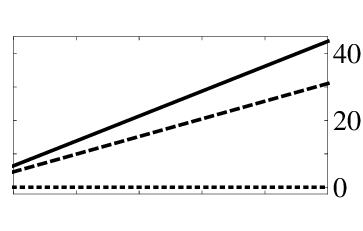}}
\begin{picture}(0,0)(133,115)
\put(-120,168){$\boldsymbol E_N$}
\put(-15,132){\gray (e)}
\put(103,132){\gray (f)}
\put(-78,115){10}
\put(-38,115){30}
\put(-2,115){50}
\put(-25,107){$\boldsymbol n_B$}
\put(16,115){4}
\put(55,115){12}
\put(96,115){20}
\put(76,107){$\boldsymbol l_B$}
\end{picture}
\caption{(Left panels) From top to bottom: fidelity $F_I(\beta)$,
mutual information $I(\beta)$, and negativity of entanglement
$E_N(\beta)$ for 1D systems as a function of the block size $n_B$
(the system is composed of $n_S=400$ oscillators and the coupling
is $c=0.4999$). (Right panels) From top to bottom: $F_I(\beta)$,
$I(\beta)$, and $E_N(\beta)$ for 2D systems as a function of $l_B$
($n_B=l_B\times l_B$, $n_S=4n_B$, $c=0.24$). In all panels, the
inverse temperature is $\beta=1,5,10$ for the dotted, dashed, and
solid lines, respectively. Note that all the plotted quantities
follow an area law.} \label{1D2D}
\end{figure}

We have now introduced all the ingredients needed to study the
behaviour of the intensive fidelity \eqref{intfid}. The dependence
of $F_I(\beta)$ with the block size is plotted in
Figs.~\ref{F1_nB} and 2(b), for 1D and 2D systems respectively.
Notice first that there are parameters for which $F_I(\beta)$ is
very close to one: the block is there well approximated by a
thermal state even when refined measurements are available. This
means that, in these cases, the thermodynamic paradigm of local
intensive temperature applies, even beyond the standard
thermodynamics setting. However, there are also parameters for
which the intensive fidelity is significantly smaller than one.
Actually, an interesting behaviour emerges: the fidelity either
stays constant (1D) or drops (2D) as the block gets larger,
contrary to what considerations on the energy balance between
subparts of the system would suggest.
In particular, for 2D systems the fidelity may drop to zero in the
limit of macroscopically large blocks, despite the interactions at
the boundary of the block being negligible with respect to the
ones in its bulk. That is, an observer able to master all possible
measurements will detect the failure of temperature intensiveness
easier for larger systems.

Clearly, the energy balance between subparts of the system cannot
be responsible of this counter-intuitive result. In particular,
the standard thermodynamic argument to show the intensiveness of
temperature (or, equivalently, the extensiveness of the entropy)
invokes the fact that the interaction at the border between $B$
and $R$ becomes negligible, for large block sizes, with respect to
the energy of $B$ and $R$ \cite{entropy}. However, in the
refined-measurement setting adopted here, this border effect may
still be---and in fact happens to be---relevant. More
specifically, correlations are responsible of this result. Recall
that in the considered systems, the correlations between $B$ and
$R$ follow an \textit{area law}:  when increasing $B$,
correlations saturate for 1D systems---since the boundary between
$B$ and $R$ stays constant---while they change linearly with $l_B$
for 2D systems. This is precisely the same behaviour observed for
the intensive fidelity. The crucial point identified here is that
$F_I(\beta)$ is such a highly sensitive quantity that detects
boundary effects \cite{note_Fsensitivity}.

\begin{figure}
\subfigure{\label{FBCS1}
\includegraphics[width=4.15cm]{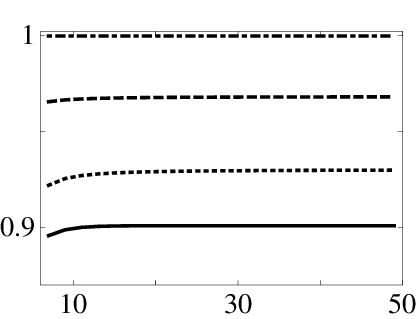}}
\subfigure{\label{FBCS2}
\includegraphics[width=4.15cm]{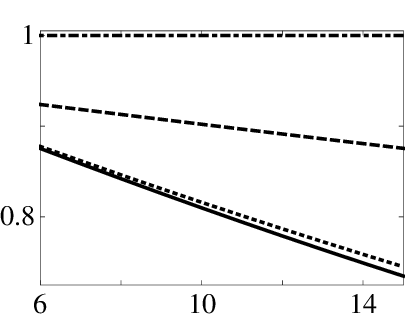}}
\begin{picture}(3,3)(117,52)
\rput{90}(0,4.4){\textbf{Fidelity}}
\put(85,60){$\boldsymbol n_B$}
\put(200,60){$\boldsymbol l_B$}
\put(85,135){\scriptsize Cores}
\put(53,115){\scriptsize Shells (1 layer)}
\put(50,95){\scriptsize Shells (2 layers)}
\put(82,78){\scriptsize Blocks}
\put(14,80){\gray (a)}
\put(213,135){\scriptsize Cores}
\put(183,123){\scriptsize Shells (1 layer)}
\put(180,100){\scriptsize Shells (2 layers)}
\put(185,78){\scriptsize Blocks}
\put(137,80){\gray (b)}
\end{picture}
\vskip -0.4cm
\caption{Right and left panels correspond to 1D and 2D systems respectively. From bottom to top: fidelities between the distinguished block $\rho_B(\beta)$ and the reference thermal state $\Omega'_B(\beta)$, their shells (composed of two and one layer), and their cores (with $n_C=n_B-2$ for 1D and $l_C=l_B-2$ for 2D). We set $\beta=10$ in both panels and other parameters as in Fig.~\ref{1D2D}.}
\label{fbcsaf}
\end{figure}
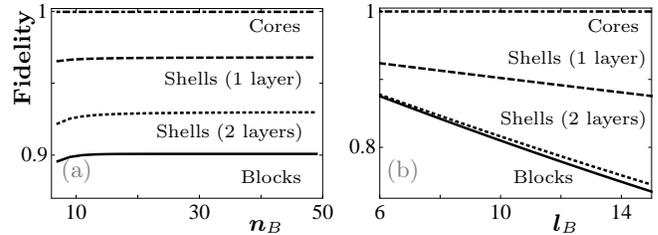

In order to confirm the foregoing intuition, we compare the core of the block and the core of the reference thermal state by tracing out few boundary layers of the two states (see Fig.~\ref{lattice} for 2D systems with core composed of $n_C=l_C \times l_C$ oscillators). The resulting fidelity is indistinguishable from one, as shown in Fig.~\ref{fbcsaf}. We can also see that all the deviation from intensiveness resides in the shell surrounding the core. This observation has two operational consequences. First, the actual state and the reference thermal state become more indistinguishable when increasing the size for standard coarse-grained measurements. Consider for instance the internal energy $U=\langle H \rangle$. For large systems, both $\Tr(\rho_B H)$ and $\Tr(\Omega_B H)$ are approximately equal and given by the value of $U$ at the core.
Clearly, the same reasoning can be applied to any observable consisting of averages of local observables. Second, it is possible to define an effective thermal state for any size and observable. It suffices to consider a thermal state for a slightly bigger system, $B+\varepsilon$, and trace out the shell $\varepsilon$ to take into account boundary effects. That is, the resulting thermal state for the block reads
\begin{equation}
    \tilde\Omega_B(\beta)=\Tr_\varepsilon\Omega_{B+\varepsilon}(\beta) .
\end{equation}
As shown in Fig.~\ref{fbcsaf}, the fidelity between the actual state and this effective thermal state is always very close to one. This approximation works remarkably well even for small sizes, near criticality and for all observables.

\section{Quantum and classical correlations}
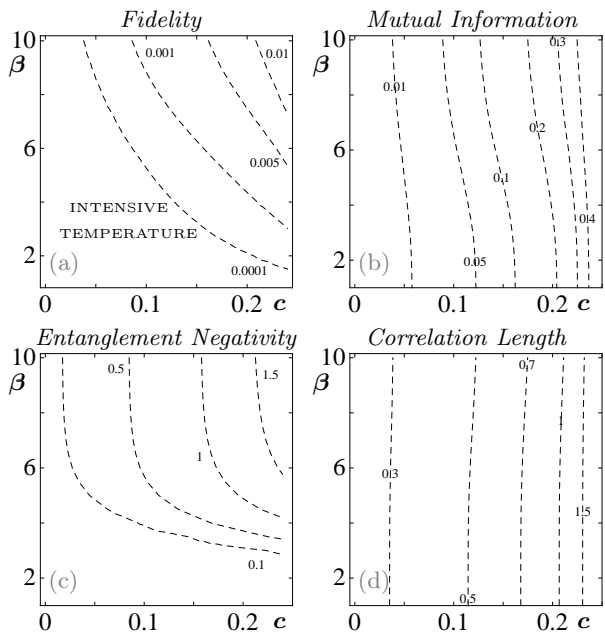
\begin{figure}
\subfigure{\label{aF_cb}
\includegraphics[width=3.9cm]{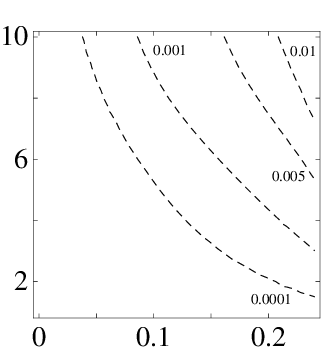}}
\subfigure{\label{aMI_cb}
\includegraphics[width=3.9cm]{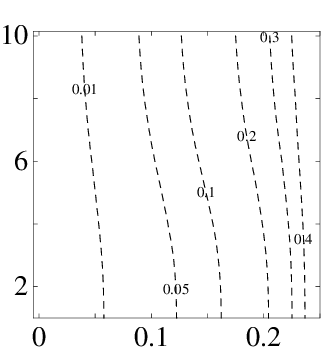}}
\begin{picture}(3,3)(117,52)
\put(-73,162){\it Fidelity}
\put(-115,145){$\boldsymbol \beta$}
\put(-15,55){$\boldsymbol c$}
\put(-100,70){\gray (a)}
\put(-92,92){\tiny INTENSIVE}
\put(-96,82){\tiny TEMPERATURE}
\put(20,162){\it Mutual Information}
\put(0,145){$\boldsymbol \beta$}
\put(100,55){$\boldsymbol c$}
\put(17,70){\gray (b)}
\end{picture}\\
\vskip -0.1cm
\subfigure{\label{aEN_cb}
\includegraphics[width=3.9cm]{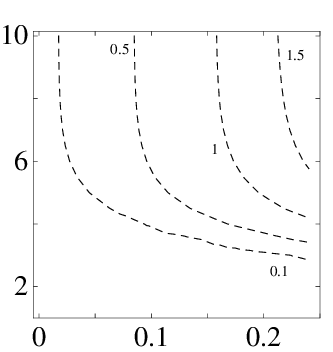}}
\subfigure{\label{CL_cb}
\includegraphics[width=3.9cm]{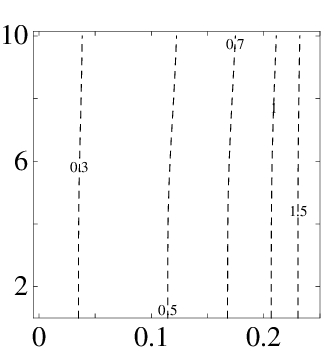}}
\begin{picture}(3,3)(117,52)
\put(-105,162){\it Entanglement Negativity}
\put(-115,145){$\boldsymbol \beta$}
\put(-15,55){$\boldsymbol c$}
\put(-100,70){\gray (c)}
\put(20,162){\it Correlation Length}
\put(0,145){$\boldsymbol \beta$}
\put(100,55){$\boldsymbol c$}
\put(17,70){\gray (d)}
\end{picture}
\caption{Contour plots of the following quantities: (a) slope $\alpha_F$ of the fidelity; (b) slope $\alpha_I$ of the mutual information; (c) slope $\alpha_E$ of the negativity of entanglement; (d) two-point correlation length (see text). Panel (a) shows that the temperature is locally intensive for high temperatures or weak coupling. The quantities in the right panels (b,d)---insensitive to the distinction between quantum and classical correlations---are almost temperature independent, thus failing to identify regions where temperature is not intensive. On the other hand, the negativity of entanglement---an indicator of purely quantum correlations---is effective in singling out the loss of intensiveness. The system is a 2D lattice with $n_S=400$.}
\end{figure}

The previous analysis points out the key role correlations play
for an intensive behaviour of the temperature. Note that, as the
system is mixed, both classical and quantum correlations can
coexist. Our last goal is to identify which correlations are
responsible of the breakdown of temperature intensiveness.

The analysis on classical harmonic systems already suggested that
classical correlations should play no role. Another intuition
pointing at the relevance of quantum correlations stems from the
following argument. Consider a generic quantum system and two
extreme cases, zero and infinite temperature. In the latter case,
the system is factorized and maximally mixed, as well as any block
of it: the temperature is intensive. For zero temperature,
instead, two situations might occur (provided the ground state is
pure; \ie, any possible degeneracy has been broken): either the
ground state is entangled, thus enforcing non-zero entropy and
non-zero temperature for any block, or it is factorized and any
block is pure. Thus for these two extreme cases we have that only
in the absence of entanglement the temperature is intensive.

To address the origin of the intensiveness breakdown
quantitatively, let us consider both the total correlations, given
by the mutual information $I(\beta)$, and a form of genuinely
quantum correlations, given by the entanglement negativity
$E_N(\beta)$ \cite{ECP}. We have calculated $I(\beta)$ and
$E_N(\beta)$ in the $B$ vs.~$R$ partition. In Figs. 2(c), 2(d),
\ref{MI2_LB}, and \ref{EN2_LB} we can see that both quantities
follow an area law, as expected~\cite{ECP}. There is however an
important difference: while $I(\beta)$ takes finite values in the
limit of infinite temperature~\cite{CEP+}, $E_N(\beta)$ drops to
zero. That is, while total correlations are present even in
situations where the intensive fidelity is nearly one, our
calculations indicate a much better agreement between the presence
of entanglement and intensiveness breakdown.

Finally, in order to get a better picture of this connection, we
introduce size-independent quantities. We focus here on 2D
systems, but similar results hold true for 1D. Consider the slope
$\alpha_F$ of the fidelity decay as a function of the block size:
$F_I(\beta)\approx\alpha_F l_B$ (in the case of 1D systems one
should consider the fidelity saturation value). This slope gives
an assessment of the intensiveness breakdown, since the smaller it
is the more intensive the temperature. This allows us to depict a
complete ``intensiveness phase diagram'' in the $c\!-\!\beta$
plane, by plotting the contours of $\alpha_F$ [see
Fig.~\ref{aF_cb}]. We recognize two regions: for low temperatures
and strong coupling the system behaves non-intensively, whereas
intensiveness is recovered in the region of high temperatures or
weak coupling. A crossover between these two behaviours appears
for intermediate $\beta$ and $c$.

Following the same procedure for mutual information and
entanglement, we introduce the slopes $\alpha_I$ and $\alpha_E$,
respectively, and plot their ``phase diagrams'' (see
Figs.~\ref{aMI_cb} and \ref{aEN_cb}). We can see that total
correlations play no role in the breakdown of intensiveness,
showing no relation with the regions individuated in
Fig.~\ref{aF_cb}. On the other hand, the relation between
intensiveness and entanglement is enforced, pointing out the role
of genuine quantum correlations in detecting the intensive region.
In fact, similar considerations hold true also if we consider
other size-independent quantities. For example, the correlation
length $\xi$ (given by $\langle q_i
q_{i+r}\rangle\propto\exp\{-r/\xi\}$)---extensively studied in
condensed matter systems but unable to distinguish quantum from
classical correlations---fails at detecting the intensive region
and shows a similar behaviour as the mutual information [see
Fig.~\ref{CL_cb}]. Although we cannot claim that the intensiveness
breakdown \textit{strictly} depends on the block negativity, as
the contour plots of Figs.~\ref{aF_cb} and \ref{aEN_cb} do not
strictly coincide, temperature ceases to be intensive in the
regions where entanglement is significantly present. Vice-versa,
without quantum correlations the thermodynamic paradigm of
intensive temperature applies.

In conclusion, we have tested the concept of intensive temperature
when the standard thermodynamics prescription of coarse
measurement is relaxed and more refined measurements are at
disposal. For systems composed of quantum oscillators, we have
seen that the thermodynamic paradigm of local intensive
temperature applies also to this setting whenever the presence of
entanglement in the system is negligible. This extends the concept
of temperature, and assesses its limits of validity, to a scenario
beyond the standard one and that might be relevant for future
technologies at mesoscopic and nanoscopic scale.

{\it Acknowledgements:} We thank L. Masanes for insightful
discussion. This work was supported by the European FP7 COMPAS project, ERC
Starting grant PERCENT, and the Marie Curie IEF No 255624, the Spanish
FIS2010-14830 project and Generalitat de Catalunya.



\begin{thebibliography}{99}
\bibitem{note_thermalization} Let us stress that we consider systems that have already undergone thermalization, possibly via weak coupling with a thermal environment. We are not addressing the thermalization dynamics leading to the thermal state of the oscillators and our approach is static in this sense.
\bibitem{note_integrable} The generalization of our analysis to non-integrable systems is outside the scope of the present investigation and could be the subject of future studies.
\end{thebibliography}
\end{document}